\documentstyle[12pt]{article}
\input tcilatex

\begin{document}

\begin{titlepage}
\date{(December 31, 1999; version 4: May 8, 2000)}
\title{ Locally Anisotropic Black Holes\\ in Einstein Gravity }
\author{ Sergiu I.\ Vacaru \thanks{\copyright\ Sergiu I. Vacaru,\ e-mail:
vacaru@lises.asm.md}  \\[10pt]
  \small Institute of Applied Physics,  Academy of Sciences,\\
  \small Academy str. 5,\  Chi\c sin\v au MD2028, \\
  \small Republic of Moldova     }
\maketitle
PACS: 04.20.Jb, 04.70.Bw \qquad {\bf gr--qc/0001020}
\vskip 2 cm
\thispagestyle{empty}
\begin{abstract}
\baselineskip18pt
 By applying the method of moving frames  modelling one and two dimensional
  local anisotropies we construct  new solutions of Einstein equations
 on  pseudo--Riemannian spacetimes. The first class of solutions
 describes non--trivial deformations of static spherically symmetric
 black holes to locally anisotropic ones which have elliptic (in three
 dimensions) and  ellipsoidal, toroidal and  elliptic and another forms
 of cylinder  symmetries (in four dimensions). The  second class consists
 from  black holes with oscillating  elliptic horizons.
 \end{abstract}
\end{titlepage}
\newpage

\section{Introduction}

In recent years, there has been great interest in investigation of
gravitational models with anisotropies and applications in modern cosmology
and astrophysics. There are possible locally anisotropic inflational and
black hole like solutions of Einstein equations in the framework of
so--called generalized Finsler--Kaluza--Klein models \cite{v1} and in
low--energy locally anisotropic limits of (super) string theories \cite{v2}.

In this paper we shall restrict ourselves to a more limited problem of
definition of black hole solutions with local anisotropy in the framework of
the Einstein theory (in three and four dimensions). Our purpose is to
construct solutions of gravitational field equations by imposing symmetries
differing in appearance from the static spherical one (which uniquely
results in the Schwarzschild solution) and search for solutions with
configurations of event horizons like rotation ellipsoids, torus and
ellipsoidal and cylinders. We shall proof that there are possible elliptic
oscillations in time of horizons.

In order to simplify the procedure of solution and investigate more deeply
the physical implications of general relativistic models with local
anisotropy we shall transfer our analysis with respect to anholonomic frames
which are equivalently characterized by nonlinear connection (N--connection)
structures \cite{barthel,cartan,ma,v1,v2}. This geometric approach is very
useful for construction of metrics with prescribed symmetries of horizons
and definition of conditions when such type black hole like solutions could
be selected from an integral variety of the Einstein field equations with a
corresponding energy--momentum tensor. We argue that, in general, the
symmetries of solutions are not completely determined by the field equations
and coordinate conditions but there are also required some physical
motivations for choosing of corresponding classes of systems of reference
(prescribed type of local anisotropy and symmetries of horizons) with
respect to which the 'picture' of interactions, symmetries and conservation
laws is drawn in the simplest form.

The paper is organized as follows: In section 2 we introduce metrics and
anholonomic frames with local anisotropies admitting equivalent
N--connection structures. We write down the Einstein equations with respect
to such locally anisotropic frames. In section 3 we analyze the general
properties of the system of gravitational field equations for an ansatz for
metrics with local anisotropy. In section 4 we generalize the three
dimensional static black hole solution to the case with elliptic horizon and
proof that there are possible elliptic oscillations in time of locally
anisotropic black holes. The section 5 is devoted to four dimensional
locally anisotropic static solutions with rotation ellipsoidal, toroidal and
cylindrical like horizons and consider elliptic oscillations in time. In the
last section we make some final remarks.

\section{Anholonomic frames and N--con\-nec\-ti\-ons}

\setcounter{equation}{0}

In this section we outline the necessary results on spacetime differential
geometry \cite{haw} and anholonomic frames induced by N--connection
structures \cite{ma,v1,v2}. We examine an ansatz for locally anisotropic
(pseudo) Riemannian metrics with respect to coordinate bases and illustrate
a substantial geometric simplification and reduction of the number of
coefficients of geometric objects and field equations after linear
transforms to anholonomic bases defined by coefficients of a corresponding
N--connection. The Einstein equations are rewritten in an invariant form
with respect to such locally anisotropic bases.

Consider a class of pseudo--Riemannian metrics
$$
g=g_{\underline{\alpha }\underline{\beta }}\left( u^\varepsilon \right) ~du^{%
\underline{\alpha }}\otimes du^{\underline{\beta }}
$$
in a $n+m$ dimensional spacetime $V^{(n+m)},\
  (n=2$ and $m=1,2$), with components
\begin{equation}
\label{metric}g_{\underline{\alpha }\underline{\beta }}=\left[
\begin{array}{cc}
g_{ij}+N_i^aN_j^bh_{ab} & N_j^eh_{ae} \\
N_i^eh_{be} & h_{ab}
\end{array}
\right] ,
\end{equation}
where $g_{ij}=g_{ij}\left( u^\alpha \right) $ and $h_{ab}=h_{ab}\left(
u^\alpha \right) $ are respectively some symmetric $n\times n$ and $m\times
m $ dimensional matrices, $N_j^e=N_j^e\left( u^\beta \right) $ is a $n\times
m $ matrix, and the $n+m$ dimensional local coordinates are
 provide with general Greek indices and denoted $u^\beta
=(x^i,y^a).$  The Latin indices $i,j,k,...$ in (\ref{metric}) run values
$1,2$ and $a,b,c,..$ run values $3,4$ and we note that both type of isotropic,
 $x^i,$ and the so--called anisotropic, $y^a,$ coordinates could be space or
time like ones. We  underline indices in order to emphasize that
components are given with respect to a coordinate (holonomic) basis
\begin{equation}
\label{hb}e_{\underline{\alpha }}=\partial _{\underline{\alpha }}=\partial
/\partial u^{\underline{\alpha }}
\end{equation}
and/or its dual
\begin{equation}
\label{dhb}e^{\underline{\alpha }}=du^{\underline{\alpha }}.
\end{equation}

The class of metrics (\ref{metric}) transform into a $(n\times n)\oplus
(m\times m)$ block form
\begin{equation}
\label{dmetric}g=g_{ij}\left( u^\varepsilon \right) ~dx^i\otimes
dx^j+h_{ab}\left( u^\varepsilon \right) \left( \delta y^a\right) ^2\otimes
\left( \delta y^a\right) ^2
\end{equation}
if one chooses a frame of basis vectors
\begin{equation}
\label{dder}\delta _\alpha =\delta /\partial u^\alpha =\left( \delta
/\partial x^i=\partial _i-N_i^a\left( u^\varepsilon \right) \partial
_a,\partial _b\right) ,
\end{equation}
where $\partial _i=\partial /x^i$ and $\partial _a=\partial /\partial y^a,$
with the dual basis being
\begin{equation}
\label{ddif}\delta ^\alpha =\delta u^\alpha =\left( dx^i,\delta
y^a=dy^a+N_i^a\left( u^\varepsilon \right) dx^i\right) .
\end{equation}

The set of coefficients $N=\{N_i^a\left( u^\varepsilon \right) \}$ from (\ref
{dder}) and (\ref{ddif}) could be associated to components of a nonlinear
connection (in brief, N--connection) structure defining a local
decomposition of spacetime into $n$ isotropic directions $x^i$ and one or
two anisotropic directions $y^a.$ The global definition of N--connection is
due to W. Barthel \cite{barthel} (the rigorous mathematical definition of
N--connection is possible on the language of exact sequences of vector, or
tangent, subbundles) and this concept is largely applied in Finsler geometry
and its generalizations \cite{cartan,ma}. It was concluded \cite{v1,v2} that
N--connection structures are induced under non--trivial dynamical
compactifications of higher dimensions in (super) string and (super) gravity
theories and even in general relativity if we are dealing with anholonomic
frames.

A N--connection is characterized by its curvature, N--curvature,
\begin{equation}
\label{ncurv}\Omega _{ij}^a=\partial _iN_j^a-\partial _jN_i^a+N_i^b\partial
_bN_j^a-N_j^b\partial _bN_i^a.
\end{equation}
As a particular case we obtain a linear connection field $\Gamma
_{ib}^a\left( x^i\right) $ if $N_i^a(x^i,y^a) = \Gamma _{ib}^a\left( x^i,
y^a \right) $ \cite{ma,v1}.

For nonvanishing values of $\Omega _{ij}^a$ the basis (\ref{dder}) is
anholonomic and satisfies the conditions
$$
\delta _\alpha \delta _\beta -\delta _\beta \delta _\alpha =w_{~\alpha \beta
}^\gamma \delta _\gamma ,
$$
where the anholonomy coefficients $w_{~\alpha \beta }^\gamma $ are defined
by the components of N--connection,
\begin{eqnarray}
w_{~ij}^k & = & 0,w_{~aj}^k=0,w_{~ia}^k=0,w_{~ab}^k=0,w_{~ab}^c=0,
\nonumber\\
w_{~ij}^a & = &
-\Omega _{ij}^a,w_{~aj}^b=-\partial _aN_i^b,w_{~ia}^b=\partial _aN_i^b.
\nonumber
\end{eqnarray}

We emphasize that the elongated by N--connection operators (\ref{dder}) and (%
\ref{ddif}) must be used, respectively, instead of local operators of
partial derivation (\ref{hb}) and differentials (\ref{dhb}) if some
differential calculations are performed with respect to any anholonomic
bases locally adapted to a fixed N--connection structure (in brief, we shall
call such local frames as la--bases or la--frames, where, in brief,
 la-- is from locally anisotropic).

The torsion, $T\left( \delta _\gamma ,\delta _\beta \right) =T_{~\beta
\gamma }^\alpha \delta _\alpha ,$ and curvature, $R\left( \delta _\tau
,\delta _\gamma \right) \delta _\beta =R_{\beta ~\gamma \tau }^{~\alpha
}\delta _\alpha ,$ tensors of a linear connection $\Gamma _{~\beta \gamma
}^\alpha $ are introduced in a usual manner and, respectively, have the
components
\begin{equation}
\label{torsion}T_{~\beta \gamma }^\alpha =\Gamma _{~\beta \gamma }^\alpha
-\Gamma _{~\gamma \beta }^\alpha +w_{~\beta \gamma }^\alpha
\end{equation}
and
\begin{equation}
\label{curvature}R_{\beta ~\gamma \tau }^{~\alpha }=\delta _\tau \Gamma
_{~\beta \gamma }^\alpha -\delta _\gamma \Gamma _{~\beta \delta }^\alpha
+\Gamma _{~\beta \gamma }^\varphi \Gamma _{~\varphi \tau }^\alpha -\Gamma
_{~\beta \tau }^\varphi \Gamma _{~\varphi \gamma }^\alpha +\Gamma _{~\beta
\varphi }^\alpha w_{~\gamma \tau }^\varphi .
\end{equation}

The Ricci tensor is defined
\begin{equation}
\label{ricci}R_{\beta \gamma }=R_{\beta ~\gamma \alpha }^{~\alpha }
\end{equation}
and the scalar curvature is
\begin{equation}
\label{scalarcurvature}R=g^{\beta \gamma }R_{\beta \gamma } .
\end{equation}

The Einstein equations with respect to a la--basis (\ref{ddif}) are written
\begin{equation}
\label{einsteq1}R_{\beta \gamma }-\frac R2g_{\beta \gamma }=k\Upsilon
_{\beta \gamma },
\end{equation}
where the energy--momentum d--tensor $\Upsilon _{\beta \gamma }$ includes
the cosmological constant terms and possible contributions of torsion (\ref
{torsion}) and matter and $k$ is the coupling constant. For a symmetric
linear connection the torsion field can be considered as induced by the
anholonomy coefficients. For dynamical torsions there are necessary
additional field equations, see, for instance, the case of locally
anisotropic gauge like theories \cite{vg}.

The geometrical objects with respect to a la--bases are distinguished by the
corresponding N--connection structure and called (in brief) d--tensors,
d--metrics (\ref{dmetric}), linear d--connections and so on \cite{ma,v1,v2}.

A linear d--connection $D$ on a spacetime $V,$
$$
D_{\delta _\gamma }\delta _\beta =\Gamma _{~\beta \gamma }^\alpha \left(
x^k,y^a\right) \delta _\alpha ,
$$
is parametrized by non--trivial horizontal (isotropic) -- vertical
(anisotropic), in brief, h--v--components,
\begin{equation}
\label{dcon}\Gamma _{~\beta \gamma }^\alpha =\left(
L_{~jk}^i,L_{~bk}^a,C_{~jc}^i,C_{~bc}^a\right) .
\end{equation}
Some d--connection and d--metric structures are compatible if there are
satisfied the conditions
$$
D_\alpha g_{\beta \gamma }=0.
$$
For instance, the canonical compatible d--connection
$$
^c\Gamma _{~\beta \gamma }^\alpha =\left(
^cL_{~jk}^i,^cL_{~bk}^a,^cC_{~jc}^i,^cC_{~bc}^a\right)
$$
is defined by the coefficients of d--metric (\ref{dmetric}), $g_{ij}\left(
x^i,y^a\right) $ and $h_{ab}\left( x^i,y^a\right) ,$ and of N--connection, $%
N_i^a=N_i^a\left( x^i,y^b\right) ,$
\begin{eqnarray}
^cL_{~jk}^i & = & \frac 12g^{in}\left( \delta _kg_{nj}+\delta _jg_{nk}-\delta
_ng_{jk}\right) , \label{cdcon} \\
^cL_{~bk}^a & = & \partial _bN_k^a+\frac 12h^{ac}\left( \delta
_kh_{bc}-h_{dc}\partial _bN_i^d-h_{db}\partial _cN_i^d\right) ,
\nonumber \\
^cC_{~jc}^i & = & \frac 12g^{ik}\partial _cg_{jk}, \nonumber \\
^cC_{~bc}^a & = & \frac 12h^{ad}\left( \partial _ch_{db}+\partial
_bh_{dc}-\partial _dh_{bc}\right) .  \nonumber
\end{eqnarray}
The coefficients of the canonical d--connection generalize with respect to
la--bases the well known Cristoffel symbols.

For a d--connection (\ref{dcon}) we can compute the non--trivial components
of d--torsion (\ref{torsion})
\begin{eqnarray}
T_{.jk}^i & = & T_{jk}^i=L_{jk}^i-L_{kj}^i,\quad
T_{ja}^i=C_{.ja}^i,T_{aj}^i=-C_{ja}^i, \nonumber \\
T_{.ja}^i & = & 0,\quad T_{.bc}^a=S_{.bc}^a=C_{bc}^a-C_{cb}^a,
\label{dtors} \\
T_{.ij}^a & = &
-\Omega _{ij}^a,\quad T_{.bi}^a= \partial _b  N_i^a
-L_{.bj}^a,\quad T_{.ib}^a=-T_{.bi}^a. \nonumber
\end{eqnarray}

In a similar manner, putting non--vanishing coefficients (\ref{dcon}) into
the formula for curvature (\ref{curvature}), we can compute the coefficients
of d--curvature
$$
R\left( \delta _\tau ,\delta _\gamma \right) \delta _\beta = R_{\beta
~\gamma\tau }^{~\alpha }\delta _\alpha ,%
$$
split into h--, v--invariant components,
\begin{eqnarray}
R_{h.jk}^{.i} & = & \delta _kL_{.hj}^i-\delta_jL_{.hk}^i
 +  L_{.hj}^mL_{mk}^i-L_{.hk}^mL_{mj}^i-C_{.ha}^i\Omega _{.jk}^a,
\nonumber \\
R_{b.jk}^{.a} & = & \delta _kL_{.bj}^a-\delta_jL_{.bk}^a
  +  L_{.bj}^cL_{.ck}^a-L_{.bk}^cL_{.cj}^a-C_{.bc}^a\Omega _{.jk}^c,
\nonumber \\
P_{j.ka}^{.i} & = & \partial _kL_{.jk}^i +C_{.jb}^iT_{.ka}^b
 -  ( \partial _kC_{.ja}^i+L_{.lk}^iC_{.ja}^l -
L_{.jk}^lC_{.la}^i-L_{.ak}^cC_{.jc}^i ), \nonumber \\
P_{b.ka}^{.c} & = & \partial _aL_{.bk}^c +C_{.bd}^cT_{.ka}^d
 - ( \partial _kC_{.ba}^c+L_{.dk}^{c\,}C_{.ba}^d
- L_{.bk}^dC_{.da}^c-L_{.ak}^dC_{.bd}^c ) \nonumber \\
S_{j.bc}^{.i} & = & \partial _cC_{.jb}^i-\partial _bC_{.jc}^i
 +  C_{.jb}^hC_{.hc}^i-C_{.jc}^hC_{hb}^i, \nonumber \\
S_{b.cd}^{.a} & = &\partial _dC_{.bc}^a-\partial
_cC_{.bd}^a+C_{.bc}^eC_{.ed}^a-C_{.bd}^eC_{.ec}^a. \nonumber
\end{eqnarray}

The components of the Ricci tensor (\ref{ricci}) with respect to locally
adapted frames (\ref{dder}) and (\ref{ddif}) (in this case, d--tensor) are
as follows:
\begin{eqnarray}
R_{ij} & = & R_{i.jk}^{.k},\quad
 R_{ia}=-^2P_{ia}=-P_{i.ka}^{.k},\label{dricci} \\
R_{ai} &= & ^1P_{ai}=P_{a.ib}^{.b},\quad R_{ab}=S_{a.bc}^{.c}. \nonumber
\end{eqnarray}

We point out that because, in general, $^1P_{ai}\neq ~^2P_{ia}$ the Ricci
d--tensor is non symmetric. This is a consequence of anholonomy of la--bases.

Having defined a d-metric of type (\ref{dmetric}) on spacetime $V$ we can
compute the scalar curvature (\ref{scalarcurvature}) of a d-connection $D,$%
\begin{equation}
\label{dscalar}{\overleftarrow{R}}=G^{\alpha \beta }R_{\alpha \beta }=%
\widehat{R}+S,
\end{equation}
where $\widehat{R}=g^{ij}R_{ij}$ and $S=h^{ab}S_{ab}.$

Now, by introducing the values of (\ref{dricci}) and (\ref{dscalar}) into
equations (\ref{einsteq1}), the Einstein equations with respect to a
la--basis seen to be%
\begin{eqnarray}
R_{ij}-\frac 12\left( \widehat{R}+S\right) g_{ij} & = &
k\Upsilon _{ij}, \label{einsteq2} \\
S_{ab}-\frac 12\left( \widehat{R}+S\right) h_{ab} & = & k\Upsilon _{ab},
 \nonumber \\
^1P_{ai} & = & k\Upsilon _{ai}, \nonumber \\
^2P_{ia} & = & -k\Upsilon _{ia}, \nonumber
\end{eqnarray}
where $\Upsilon _{ij},\Upsilon _{ab},\Upsilon _{ai}$ and $\Upsilon _{ia}$
are the components of the energy--momentum d--tensor field $\Upsilon _{\beta
\gamma }$ (which includes possible cosmological constants, contributions of
anholonomy d--torsions (\ref{dtors}) and matter) and $k$ is the coupling
constant. For simplicity, we omitted the upper left index $c$ pointing that
for the Einstein theory the Ricci d--tensor and curvature scalar should be
computed by applying the coefficients of canonical d--connection (\ref{cdcon}%
).

\section{An ansatz for la--metrics}

\setcounter{equation}{0}

Let us consider a four dimensional (in brief, 4D) spacetime $V^{(2+2)}$
 (with two isotropic plus two anisotropic local coordinates)
 provided with a metric  (\ref{metric}) 
 (of  signature (-,+,+,+), or  (+,+,+,-), (+,+,-,+)) 
 parametrized by a symmetric matrix of type
\begin{equation}
\label{ansatz2}\left[
\begin{array}{cccc}
g_1+q_1{}^2h_3+n_1{}^2h_4 & 0 & q_1h_3 & n_1h_4 \\
0 & g_2+q_2{}^2h_3+n_2{}^2h_4 & q_2h_3 & n_2h_4 \\
q_1h_3 & q_2h_3 & h_3 & 0 \\
n_1h_4 & n_2h_4 & 0 & h_4
\end{array}
\right]
\end{equation}
with components being some functions
$$
g_i=g_i(x^j),q_i=q_i(x^j,z),n_i=n_i(x^j,z),h_a=h_a(x^j,z)%
$$
of necessary smoothly class. With respect to a la--basis (\ref{ddif}) this
ansatz results in diagonal $2\times 2$ h-- and v--metrics for a d--metric (%
\ref{ddif}) (for simplicity, we shall consider only diagonal 2D
nondegenerated metrics  because for such dimensions every symmetric
matrix can be diagonalized).

An equivalent diagonal d--metric (\ref{dmetric}) is obtained for the
associated N--connection with coefficients being functions on three
coordinates $(x^i,z),$%
\begin{eqnarray}
N_1^3&=&q_1(x^i,z),\ N_2^3=q_2(x^i,z), \label{ncoef} \\
N_1^4&=&n_1(x^i,z),\ N_2^4=n_2(x^i,z). \nonumber
\end{eqnarray}
For simplicity, we shall use brief denotations of partial derivatives, like $%
\dot a$$=\partial a/\partial x^1,a^{\prime }=\partial a/\partial x^2,$ $%
a^{*}=\partial a/\partial z$ $\dot a^{\prime }$$=\partial ^2a/\partial
x^1\partial x^2,$ $a^{**}=\partial ^2a/\partial z\partial z.$

The non--trivial components of the Ricci d--tensor (\ref{dricci}) ( for the
ansatz (\ref{ansatz2})) when $R_1^1 = R_2^2$ and $S_3^3 = S_4^4,$ are
computed
\begin{eqnarray} \label{ricci1}
R_1^1&=&\frac 1{2g_1g_2}
[-(g_1^{^{\prime \prime }}+{\ddot g}_2)+\frac 1{2g_2}\left( {\dot g}%
_2^2+g_1^{\prime }g_2^{\prime }\right) +
\frac 1{2g_1}\left( g_1^{\prime \ 2}+%
\dot g_1\dot g_2\right) ], \\
\label{ricci2}
S_3^3&=&\frac 1{h_3h_4}[-h_4^{**}+\frac 1{2h_4}(h_4^{*})^2+%
\frac 1{2h_3}h_3^{*}h_4^{*}], \\
 &{} & \nonumber \\
 \label{ricci3}
P_{3i}&=&\frac{q_i}2[\left( \frac{h_3^{*}}{h_3}\right) ^2-
\frac{h_3^{**}}{h_3}+ \frac{h_4^{*}}{2h_4^{\ 2}}-
\frac{h_3^{*}h_4^{*}}{2h_3h_4}]  \\
 &{}&
+\frac 1{2h_4}[\frac{\dot h_4}{2h_4}h_4^{*}-
\dot h_4^{*}+\frac{\dot h_3}{2h_3}h_4^{*}], \nonumber  \\
 &{} & \nonumber \\
 P_{4i}&=&-\frac{h_4}{2h_3}n_i^{**}. \label{ricci4}
\end{eqnarray}

The curvature scalar $\overleftarrow{R}$ (\ref{dscalar}) is defined by two
non-trivial components $\widehat{R}=2R_1^1$ and $S=2S_3^3.$

The system of Einstein equations (\ref{einsteq2}) transforms into
\begin{eqnarray}
R_1^1&=&-\kappa \Upsilon _3^3=-\kappa \Upsilon _4^4,
\label{einsteq3a} \\
S_3^3&=&-\kappa \Upsilon _1^1=-\kappa \Upsilon _2^2, \label{einsteq3b}\\
P_{3i}&=& \kappa \Upsilon _{3i}, \label{einsteq3c} \\
P_{4i}&=& \kappa \Upsilon _{4i}, \label{einsteq3d}
\end{eqnarray}
where the values of $R_1^1,S_3^3,P_{ai},$ are taken respectively from (\ref
{ricci1}), (\ref{ricci2}), (\ref{ricci3}), (\ref{ricci4}).

We note that we can define the N--coefficients (\ref{ncoef}), $q_i(x^k,z)$
and $n_i(x^k,z),$ by solving the equations (\ref{einsteq3c}) and (\ref
{einsteq3d}) if the functions $h_i(x^k,z)$ are known as solutions of the
equations (\ref{einsteq3b}).

Let us analyze the basic properties of equations (\ref{einsteq3b})--(\ref
{einsteq3d}) (the h--equa\-ti\-ons will be considered for 3D and 4D
in the next sections). The v--component of the Einstein equations (%
\ref{einsteq3a})
\begin{equation}
\label{heq}\frac{\partial ^2h_4}{\partial z^2} - \frac 1{2h_4}\left( \frac{%
\partial h_4}{\partial z}\right) ^2 -\frac 1{2h_3}\left( \frac{\partial h_3}{%
\partial z}\right) \left( \frac{\partial h_4}{\partial z}\right) - \frac
\kappa 2\Upsilon _1h_3h_4=0 \nonumber
\end{equation}
(here we write down the partial derivatives on $z$ in explicit form) follows
from (\ref{ricci2}) and (\ref{einsteq3b}) and relates some first and second
order partial on $z$ derivatives of diagonal components $h_a(x^i,z)$ of a
v--metric with a source $\kappa \Upsilon _1(x^i,z)=\kappa \Upsilon
_1^1=\kappa \Upsilon _2^2$ in the h--subspace. We can consider as unknown
the function $h_3(x^i,z)$ (or, inversely, $h_4(x^i,z))$ for some compatible
values of $h_4(x^i,z)$ (or $h_3(x^i,z))$ and source $\Upsilon _1(x^i,z).$

By introducing a new variable $\beta =h_4^{*}/h_4$ the equation (\ref{heq})
transforms into
\begin{equation}
\label{heq1}\beta ^{*}+\frac 12\beta ^2-\frac{\beta h_3^{*}}{2h_3}-2\kappa
\Upsilon _1h_3=0
\end{equation}
which relates two functions $\beta \left( x^i,z\right) $ and $h_3\left(
x^i,z\right) .$ There are two possibilities: 1) to define $\beta $ (i. e. $%
h_4)$ when $\kappa \Upsilon _1$ and $h_3$ are prescribed and, inversely 2)
to find $h_3$ for given $\kappa \Upsilon _1$ and $h_4$ (i. e. $\beta );$ in
both cases one considers only ''*'' derivatives on $z$--variable
(coordinates $x^i$ are treated as parameters).

\begin{enumerate}
\item  In the first case the explicit solutions of (\ref{heq1}) have to be
constructed by using the integral varieties of the general Riccati equation
\cite{kamke} which by a corresponding redefinition of variables, $%
z\rightarrow z\left( \varsigma \right) $ and $\beta \left( z\right)
\rightarrow \eta \left( \varsigma \right) $ (for simplicity, we omit here
the dependencies on $x^i)$ could be written in the canonical form
$$
\frac{\partial \eta }{\partial \varsigma }+\eta ^2+\Psi \left( \varsigma
\right) =0
$$
where $\Psi $ vanishes for vacuum gravitational fields. In vacuum cases the
Riccati equation reduces to a Bernoulli equation which (we can use the
former variables) for $s(z)=\beta ^{-1}$ transforms into a linear
differential (on $z)$ equation,
\begin{equation}
\label{heq1a}s^{*}+\frac{h_3^{*}}{2h_3}s-\frac 12=0.
\end{equation}

\item  In the second (inverse) case when $h_3$ is to be found for some
prescribed $\kappa \Upsilon _1$ and $\beta $ the equation (\ref{heq1}) is to
be treated as a Bernoulli type equation,
\begin{equation}
\label{heq2}h_3^{*}=-\frac{4\kappa \Upsilon _1}\beta (h_3)^2+\left( \frac{%
2\beta ^{*}}\beta +\beta \right) h_3
\end{equation}
which can be solved by standard methods. In the vacuum case the squared on $%
h_3$ term vanishes and we obtain a linear differential (on $z)$ equation.
\end{enumerate}

A particular interest presents those solutions of the equation (\ref{heq1})
which via 2D conformal transforms with a factor $\omega =\omega (x^i,z)$ are
equivalent to a diagonal h--metric on $x$--variables, i.e. one holds the
parametrization
\begin{equation}
\label{conf4d}h_3=\omega (x^i,z)\ a_3\left( x^i\right) \mbox{ and }%
h_4=\omega (x^i,z)\ a_4\left( x^i\right) ,
\end{equation}
where $a_3\left( x^i\right) $ and $a_4\left( x^i\right) $ are some arbitrary
functions (for instance, we can impose the condition that they describe some
2D soliton like or black hole solutions). In this case $\beta =\omega
^{*}/\omega $ and for $\gamma =\omega ^{-1}$ the equation (\ref{heq1})
transforms into
\begin{equation}
\label{confeq}\gamma \ \gamma ^{**}=-2\kappa \Upsilon _1a_3\left( x^i\right)
\end{equation}
with the integral variety determined by
\begin{equation}
\label{confeqsol}z=\int \frac{d\gamma }{\sqrt{\left| -4k\Upsilon
_1a_3(x^i)\ln |\gamma |+C_1(x^i)\right| }}+C_2(x^i),
\end{equation}
where it is considered that the source $\Upsilon _1$ does not depend on $z.$

Finally, we conclude that the v--metrics are defined by the integral
varieties of corresponding Riccati and/or Bernoulli equations with respect
to $z$--variables with the h--coordinates $x^i$ treated as parameters.

\section{3D black la--holes}

\setcounter{equation}{0}

Let us analyze some basic properties of 3D spacetimes $V^{(2+1)}$
 (we emphasize that in approach $(2+1)$ points to a splitting
 into two isotropic and  one anisotropic directions and not to usual
  2D space plus one time like coordinates;  in general anisotropies could
 be associate to both space and/or time like coordinates)
 provided with d--metrics of type
\begin{equation}
\label{dmetr3}\delta s^2=g_1\left( x^k\right) \left( dx^1\right)
^2+g_2\left( x^k\right) \left( dx^2\right) ^2+h_3(x^i,z)\left( \delta
z\right) ^2,
\end{equation}
where $x^k$ are 2D coordinates, $y^3=z$ is the anisotropic coordinate and
$$
\delta z=dz+N_i^3(x^k,z)dx^i.
$$
The N--connection coefficients are
\begin{equation}
\label{ncoef3}N_1^3=q_1(x^i,z),\ N_2^3=q_2(x^i,z).
\end{equation}

The non--trivial components of the Ricci d--tensor (\ref{dricci}), for the
ansatz (\ref{ansatz2}) with $h_4 =1$ and $n_i =0$, $R_1^1=R_2^2$ and $%
P_{3i}, $ are
\begin{eqnarray} \label{ricci1_3}
R_1^1&=& \frac 1{2g_1g_2} \ [-(g_1^{^{\prime \prime }}
+{\ddot g}_2)  +\frac 1{2g_2}\left( {\dot g}_2^2
+g_1^{\prime }g_2^{\prime }\right) +
\frac 1{2g_1}\left( g_1^{\prime \ 2}+%
\dot g_1\dot g_2\right) ], \\
\label{ricci3_3}
P_{3i} &=& \frac{q_i}2[\left( \frac{h_3^{*}}{h_3}\right) ^2-
\frac{h_3^{**}}{h_3} ]
\end{eqnarray}
(for 3D the component $S_3^3\equiv 0,$ see (\ref{ricci2})).

The curvature scalar $\overleftarrow{R}$ (\ref{dscalar}) is $\overleftarrow{R%
}=\widehat{R}=2R_1^1.$

The system of Einstein equations (\ref{einsteq2}) transforms into
\begin{eqnarray}
R_1^1&=&-\kappa \Upsilon _3^3,
\label{einsteq3a3} \\
P_{3i}&=& \kappa \Upsilon _{3i}, \label{einsteq3c3}
\end{eqnarray}
which is compatible for energy--momentum d--tensors with $%
\Upsilon_1^1=\Upsilon _2^2=0;$ the values of $R_1^1$ and $P_{3i}$ are taken
respectively from (\ref{ricci1_3}) and (\ref{ricci3_3}).

By using the equation (\ref{einsteq3c3}) we can define the N--coefficients (%
\ref{ncoef3}), $q_i(x^k,z),$ if the function $h_3(x^k,z)$ and the components
$\Upsilon _{3i}$ of the energy--momentum d--tensor are given. We note that
the equations (\ref{ricci3_3}) are solved for arbitrary functions $%
h_3=h_3(x^k)$ and $q_i=q_i(x^k,z)$ if $\Upsilon _{3i}=0$ and in this case
the component of d--metric $h_3(x^k)$ is not contained in the system of 3D
field equations.

\subsection{Static elliptic horizons}

Let us consider a class of 3D d-metrics which local anisotropy which are
similar to Banados--Teitelboim--Zanelli (BTZ) black holes \cite{btz}.

The d--metric is parametrized
\begin{equation}
\label{dim3}\delta s^2=g_1\left( \chi ^1,\chi ^2\right) (d\chi ^1)^2+\left(
d\chi ^2\right) ^2-h_3\left( \chi ^1,\chi ^2,t\right) \ \left( \delta
t\right) ^2,
\end{equation}
where $\chi ^1=r/r_h$ for $r_h=const,$ $\chi ^2=\theta /r_a$ if $r_a=\sqrt{%
|\kappa \Upsilon _3^3|}\neq 0$ and $\chi ^2=\theta $ if $\Upsilon _3^3=0,$ $%
y^3=z=t,$ where $t$ is the time like coordinate. The Einstein equations (\ref
{einsteq3a3}) and (\ref{einsteq3c3}) transforms respectively into
\begin{equation}
\label{hbh1a3}\frac{\partial ^2g_1}{\partial (\chi ^2)^2}-\frac
1{2g_1}\left( \frac{\partial g_1}{\partial \chi ^2}\right) ^2-2\kappa
\Upsilon _3^3g_1=0
\end{equation}
and
\begin{equation}
\label{hbh1c3}\left[ \frac 1{h_3}\frac{\partial ^2h_3}{\partial z^2}-\left(
\frac 1{h_3}\frac{\partial h_3}{\partial z}\right) ^2\right] q_i=-\kappa
\Upsilon _{3i}.
\end{equation}
By introducing new variables
\begin{equation}
\label{p-var3}p=g_1^{\prime }/g_1\mbox{ and }s=h_3^{*}/h_3
\end{equation}
where the 'prime' in this subsection denotes the partial derivative $%
\partial /\chi ^2,$ the equations (\ref{hbh1a3}) and (\ref{hbh1c3})
transform into
\begin{equation}
\label{hbh2a3}p^{\prime }+\frac{p^2}2+2\epsilon =0
\end{equation}
and
\begin{equation}
\label{hbh2c3}s^{*}q_i=\kappa \Upsilon _{3i},
\end{equation}
where the vacuum case should be parametrized for $\epsilon =0$ with $\chi
^i=x^i$ and $\epsilon =1(-1)$ for the signature $1(-1)$ of the anisotropic
coordinate.

A class of solutions of 3D Einstein equations for arbitrary $q_i=q_i(\chi
^k,t)$ and $\Upsilon _{3i}=0$ is obtained if $s=s(\chi ^i).$ After
integration of the second equation from (\ref{p-var3}), we find
\begin{equation}
\label{hbh2c3s}h_3(\chi ^k,t)=h_{3(0)}(\chi ^k)\exp \left[ s_{(0)}\left(
\chi ^k\right) t\right]
\end{equation}
as a general solution of the system (\ref{hbh2c3}) with vanishing right
part. Static solutions are stipulated by $q_i=q_i(\chi ^k)$ and $%
s_{(0)}(\chi ^k)=0.$

The integral curve of (\ref{hbh2a3}), intersecting a point $\left( \chi
_{(0)}^2,p_{(0)}\right) ,$ considered as a differential equation on $\chi ^2$
is defined by the functions \cite{kamke}%
\begin{eqnarray}
p &=&
\frac{p_{(0)}}{1+\frac{p_{(0)}}2
\left( \chi ^2-\chi _{(0)}^2\right) },\qquad   \epsilon =0; \label{eq3a} \\
p & = & \frac{p_{(0)}-2\tanh \left( \chi ^2- \chi _{(0)}^2\right) }{1+\frac{p_{(0)}}2    \tanh \left( \chi ^2-\chi _{(0)}^2\right) },\qquad  %
  \epsilon >0; \label{eq3b}  \\     p & = & \frac{p_{(0)}-2\tan \left( \chi ^2-\chi _{(0)}^2\right) } {1+\frac{p_{(0)}}2\tan \left( \chi ^2-\chi _{(0)}^2\right) },\qquad
 \epsilon <0.   \label{eq3c} %
\end{eqnarray}

Because the function $p$ depends also parametrically on variable $\chi ^1$
we must consider functions $\chi _{(0)}^2=\chi _{(0)}^2\left( \chi ^1\right)
$ and $p_{(0)}=p_{(0)}\left( \chi ^1\right) .$

For simplicity, here we elucidate the case $\epsilon <0.$ The general
formula for the nontrivial component of h--metric is to be obtained after
integration on $\chi ^1$ of (\ref{eq3c}) (see formula (\ref{p-var3}))%
$$
g_1\left( \chi ^1,\chi ^2\right) =g_{1(0)}\left( \chi ^1\right) \left\{ \sin
[\chi ^2-\chi _{(0)}^2\left( \chi ^1\right) ]+\arctan \frac 2{p_{(0)}\left(
\chi ^1\right) }\right\} ^2,
$$
for $p_{(0)}\left( \chi ^1\right) \neq 0,$ and
\begin{equation}
\label{btzlh3}g_1\left( \chi ^1,\chi ^2 \right) =g_{1(0)}\left( \chi
^1\right) \ \cos ^2[\chi ^2-\chi _{(0)}^2\left( \chi ^1\right) ]
\end{equation}
for $p_{(0)}\left( \chi ^1\right) =0,$ where $g_{1(0)}\left( \chi
^1\right),\chi _{(0)}^2\left( \chi ^1\right) $ and $p_{(0)}\left( \chi
^1\right) $ are some functions of necessary smoothness class on variable $%
\chi ^1=x^1/\sqrt{\kappa \varepsilon },$ when $\varepsilon $ is the energy
density. If we consider $\Upsilon _{3i}=0$ and a nontrivial diagonal
components of energy--momentum d--tensor, $\Upsilon _\beta ^\alpha
=diag[0,0,-\varepsilon],$ the N--connection coefficients $q_i(\chi ^i, t)$
could be arbitrary functions.

For simplicity, in our further considerations we shall apply the solution (%
\ref{btzlh3}).

The d--metric (\ref{dim3}) with the coefficients (\ref{btzlh3}) and (\ref
{hbh2c3s}) gives a general description of a class of solutions with generic
local anisotropy of the Einstein equations (\ref{einsteq2}).

Let us construct static black la--hole solutions for $s_{(0)}\left( \chi
^k\right) =0$ in (\ref{hbh2c3s}).

In order to construct an explicit la--solution we have to chose some
coefficients $h_{3(0)}(\chi ^k),g_{1(0)}\left( \chi ^1\right) $ and $\chi
_0\left( \chi ^1\right) $ from some physical considerations. For instance,
the Schwarzschild solution is selected from a general 4D metric with some
general coefficients of static, spherical symmetry by relating the radial
component of metric with the Newton gravitational potential. In this
section, we construct a locally anisotropic BTZ like solution by supposing
that it is conformally equivalent to the BTZ solution if one neglects
anisotropies on angle $\theta ),$
$$
g_{1(0)}\left( \chi ^1\right) =\left[ r\left( -M_0+\frac{r^2}{l^2}\right)
\right] ^{-2},
$$
where $M_0=const>0$ and $-1/l^2$ is a constant (which is to be considered
the cosmological from the locally isotropic limit. The time--time
coefficient of d--metric is chosen
\begin{equation}
\label{btzlva3}h_3\left( \chi ^1,\chi ^2\right) =r^{-2}\lambda _3\left( \chi
^1,\chi ^2\right) \cos ^2[\chi ^2-\chi _{(0)}^2\left( \chi ^1\right) ].
\end{equation}

If we chose in (\ref{btzlva3})
$$
\lambda _3={(-M_0+\frac{r^2}{l^2})}^2,
$$
when the constant
$$
r_h=\sqrt{M_0}l
$$
defines the radius of a circular horizon, the la--solution is conformally
equivalent, with the factor $r^{-2}\cos ^2[\chi ^2-\chi _{(0)}^2\left( \chi
^1\right) ], $ to the BTZ solution embedded into a anholonomic background
given by arbitrary functions $q_i(\chi ^i,t)$ which are defined by some
initial conditions of gravitational la--background polarization.

A more general class of la--solutions could be generated if we put, for
instance,
$$
\lambda _3\left( \chi ^1,\chi ^2\right) ={(-M}\left( \theta \right) {+\frac{%
r^2}{l^2})}^2,
$$
with
$$
{M}\left( \theta \right) =\frac{M_0}{(1+e\cos \theta )^2},
$$
where $e<1.$ This solution has a horizon, $\lambda _3=0,$ parametrized by an
ellipse
$$
r=\frac{r_h}{1+e\cos \theta }
$$
with parameter $r_h$ and eccentricity $e.$

We note that our solution with elliptic horizon was constructed for a
diagonal energy--momentum d-tensor with nontrivial energy density but
without cosmological constant. On the other hand the BTZ solution was
constructed for a generic 3D cosmological constant. There is not a
contradiction here because the la--solutions can be considered for a
d--tensor $\Upsilon _\beta ^\alpha =diag[p_1-1/l^2,p_2-1/l^2,-\varepsilon
-1/l^2]$ with $p_{1,2}=1/l^2$ and $\varepsilon _{(eff)}=\varepsilon +1/l^2$
(for $\varepsilon =const$ the last expression defines the effective constant
$r_a).$ The locally isotropic limit to the BTZ black hole could be realized
after multiplication on $r^2$ and by approximations $e\simeq 0,$ $\cos
[\theta -\theta _0\left( \chi ^1\right) ]\simeq 1$ and $q_i(x^k,t)\simeq 0.$

\subsection{Oscillating elliptic horizons}

The simplest way to construct 3D solutions of the Einstein equations with
oscillating in time horizon is to consider matter states with constant
nonvanishing values of $\Upsilon _{31}=const.$ In this case the coefficient $%
h_3$ could depend on $t$--variable. For instance, we can chose such initial
values when
\begin{equation}
\label{btzlva3osc1}h_3(\chi ^1,\theta ,t)=r^{-2}\left( -M\left( t\right) +%
\frac{r^2}{l^2}\right) \cos ^2[\theta -\theta _0\left( \chi ^1\right) ]
\end{equation}
with
$$
M=M_0\exp \left( -\widetilde{p}t\right) \sin \widetilde{\omega }t,
$$
or, for an another type of anisotropy,
\begin{equation}
\label{btzlva3osc2}h_3(\chi ^1,\theta ,t)=r^{-2}\left( -M_0+\frac{r^2}{l^2}%
\right) \cos ^2\theta \ \sin ^2[\theta -\theta _0\left( \chi ^1,t\right) ]
\end{equation}
with
$$
\cos \theta _0\left( \chi ^1,t\right) =e^{-1}\left( \frac{r_a}r\cos \omega
_1t-1\right) ,
$$
when the horizon is given parametrically, %
$$
r=\frac{r_a}{1+e\cos \theta }\cos \omega _1t,
$$
where the new constants (comparing with those from the previous subsection)
are fixed by some initial and boundary conditions as to be $\widetilde{p}>0,$
and $\widetilde{\omega }$ and $\omega _1$ are treated as some real numbers.

For a prescribed value of $h_3(\chi ^1,\theta ,t)$ with non--zero source $%
\Upsilon _{31},$ in the equation (\ref{einsteq3c3}), we obtain
\begin{equation}
\label{ncon3osc}q_1(\chi ^1,\theta ,t)=\kappa \Upsilon _{31}\left( \frac{%
\partial ^2}{\partial t^2}\ln |h_3(\chi ^1,\theta ,t)|\right) ^{-1}.
\end{equation}

A solution (\ref{dmetr3}) of the Einstein equations (\ref{einsteq3a3}) and (%
\ref{einsteq3c3}) with $g_2(\chi ^i)=1$ and $g_1(\chi ^1,\theta )$ and $%
h_3(\chi ^1,\theta ,t)$ given respectively by formulas (\ref{btzlh3}) and (%
\ref{btzlva3osc1}) describe a 3D evaporating black la--hole solution with
circular oscillating in time horizon. An another type of solution, with
elliptic oscillating in time horizon, could be obtained if we choose (\ref
{btzlva3osc2}). The non--trivial coefficient of the N--connection must be
computed following the formula (\ref{ncon3osc}).

\section{4D la--solutions}

\setcounter{equation}{0}

\subsection{Basic properties}

The purpose of this section is the construction of d--metrics which are
conformally equivalent to some la--deformations of black hole, torus and
cylinder like solutions in general relativity. We shall analyze 4D d-metrics
of type
\begin{equation}
\label{dmetr4}\delta s^2 = g_1\left( x^k\right) \left( dx^1\right) ^2+
\left(dx^2\right) ^2 + h_3(x^i,z)\left( \delta z\right) ^2 +
h_4(x^i,z)\left( \delta y^4 \right) ^2.
\end{equation}

The Einstein equations (\ref{einsteq3a}) with the Ricci h--tensor (\ref
{ricci1}) and diagonal energy momentum d--tensor transforms into
\begin{equation}
\label{hbh1}\frac{\partial ^2g_1}{\partial (x^2)^2}-\frac 1{2g_1}\left(
\frac{\partial g_1}{\partial x^2}\right) ^2-2\kappa \Upsilon _3^3g_1=0.
\end{equation}
By introducing a dimensionless coordinate, $\chi ^2=x^2/\sqrt{|\kappa
\Upsilon _3^3|},$ and the variable $p=g_1^{\prime }/g_1,$ where by 'prime'
in this section is considered the partial derivative $\partial /\chi ^2,$
the equation (\ref{hbh1}) transforms into
\begin{equation}
\label{hbh2}p^{\prime }+\frac{p^2}2+2\epsilon =0,
\end{equation}
where the vacuum case should be parametrized for $\epsilon =0$ with $\chi
^i=x^i$ and $\epsilon =1(-1).$ The equations (\ref{hbh1}) and (\ref{hbh2})
are, correspondingly, equivalent to the equations (\ref{hbh1a3}) and (\ref
{hbh2a3}) with that difference that in this section we are dealing with 4D
coefficients and values. The solutions for the h--metric are parametrized
like (\ref{eq3a}), (\ref{eq3b}), and (\ref{eq3c}) and the coefficient $%
g_1(\chi ^i)$ is given by a similar to (\ref{btzlh3}) formula (for
simplicity, here we elucidate the case $\epsilon <0)$ which for $%
p_{(0)}\left( \chi ^1\right) =0$ transforms into
\begin{equation}
\label{btzlh4}g_1\left( \chi ^1,\chi ^2\right) =g_{1(0)}\left( \chi
^1\right) \ \cos ^2[\chi ^2-\chi _{(0)}^2\left( \chi ^1\right) ],
\end{equation}
where $g_1\left( \chi ^1\right) ,\chi _{(0)}^2\left( \chi ^1\right) $ and $%
p_{(0)}\left( \chi ^1\right) $ are some functions of necessary smoothness
class on variable $\chi ^1=x^1/\sqrt{\kappa \varepsilon },$ $\varepsilon $
is the energy density. The coefficients $g_1\left( \chi ^1,\chi ^2\right) $ (%
\ref{btzlh4}) and $g_2\left( \chi ^1,\chi ^2\right) =1$ define a h--metric.
The next step is the construction of h--components of d--metrics, $%
h_a=h_a(\chi ^i,z),$ for different classes of symmetries of anisotropies.

The system of equations (\ref{einsteq3b}) with the vertical Ricci d--tensor
component (\ref{ricci2}) is satisfied by arbitrary functions
\begin{equation}
\label{hdm2var}h_3=a_3(\chi ^i)\mbox{ and }h_4=a_4(\chi ^i).
\end{equation}
For v--metrics depending on three coordinates $(\chi ^i,z)$ the
v--components of the Einstein equations transform into (\ref{heq}) which
reduces to (\ref{heq1}) for prescribed values of $h_3(\chi ^i,z),\,$ and,
inversely, to (\ref{heq2}) if $h_4(\chi ^i,z)$ is prescribed. For h--metrics
being conformally equivalent to (\ref{hdm2var}) (see transforms (\ref{conf4d}%
)) we are dealing to equations of type (\ref{confeq}) with integral
varieties (\ref{confeqsol}).

\subsection{Rotation Hypersurfaces Horizons}

We proof that there are static black hole and cylindrical like solutions of
the Einstein equations with horizons being some 3D rotation hypersurfaces.
The space components of corresponding d--metrics are conformally equivalent
to some locally anisotropic deformations of the spherical symmetric
Schwarzschild and cylindrical Weyl solutions. We note that for some classes
of solutions the local anisotropy is contained in non--perturbative
anholonomic structures.

\subsubsection{Rotation ellipsoid configuration}

There two types of rotation ellipsoids, elongated and flattened ones. We
examine both cases of such horizon configurations.

\vskip0.2cm

\paragraph{\qquad Elongated rotation ellipsoid coordinates:}

${~}$\\ ${\qquad}$
 An elongated rotation ellipsoid hypersurface is given by the formula \cite
{korn}
\begin{equation}
\label{relhor}\frac{\widetilde{x}^2+\widetilde{y}^2}{\sigma ^2-1}+\frac{%
\widetilde{z}^2}{\sigma ^2}=\widetilde{\rho }^2,
\end{equation}
where $\sigma \geq 1$ and $\widetilde{\rho }$ is similar to the radial
coordinate in the spherical symmetric case.

The space 3D coordinate system is defined%
$$
\widetilde{x}=\widetilde{\rho}\sinh u\sin v\cos \varphi ,\ \widetilde{y}=%
\widetilde{\rho}\sinh u\sin v\sin \varphi ,\ \widetilde{z}=\widetilde{\rho}%
\cosh u\cos v,
$$
where $\sigma =\cosh u,(0\leq u<\infty ,\ 0\leq v\leq \pi ,\ 0\leq \varphi
<2\pi ). $\ The hypersurface metric is
\begin{eqnarray}
g_{uu} &=& g_{vv}=\widetilde{\rho}^2\left( \sinh ^2u+\sin ^2v\right) ,
 \label{hsuf1} \\
g_{\varphi \varphi } &=&\widetilde{\rho}^2\sinh ^2u\sin ^2v.
 \nonumber
\end{eqnarray}

Let us introduce a d--metric
\begin{equation}
\label{rel1}\delta s^2 = g_1(u,v)du^2+dv^2 + h_3\left( u,v,\varphi \right)
\left( \delta t\right) ^2+h_4\left( u,v,\varphi \right) \left( \delta
\varphi \right) ^2,
\end{equation}
where $\delta t$ and $\delta \varphi $ are N--elongated differentials.

As a particular solution (\ref{btzlh4}) for the h--metric we choose the
coefficient
\begin{equation}
\label{relh1h}g_1(u,v)=\cos ^2v.
\end{equation}
The $h_3(u,v,\varphi )=h_3(u,v,\widetilde{\rho }
\left( u,v,\varphi \right) )$
is considered as
\begin{equation}
\label{relh1}h_3(u,v,\widetilde{\rho })=\frac 1{\sinh ^2u+\sin ^2v}\frac{%
\left[ 1-\frac{r_g}{4\widetilde{\rho }}\right] ^2}{\left[ 1+\frac{r_g}{4%
\widetilde{\rho }}\right] ^6}.
\end{equation}
In order to define the $h_4$ coefficient solving the Einstein equations, for
simplicity with a diagonal energy--momentum d--tensor for vanishing pressure
we must solve the equation (\ref{heq1}) which transforms into a linear
equation (\ref{heq1a}) if $\Upsilon _1=0.$ In our case $s\left( u,v,\varphi
\right) =\beta ^{-1}\left( u,v,\varphi \right) ,$ where $\beta =\left(
\partial h_4/\partial \varphi \right) /h_4,$ must be a solution of
$$
\frac{\partial s}{\partial \varphi }+\frac{\partial \ln \sqrt{\left|
h_3\right| }}{\partial \varphi }\ s=\frac 12.
$$
After two integrations (see \cite{kamke}) the general solution for $%
h_4(u,v,\varphi ),$ is
\begin{equation}
\label{relh1a}h_4(u,v,\varphi )=a_4\left( u,v\right) \exp \left[
-\int\limits_0^\varphi F(u,v,z)\ dz\right] ,
\end{equation}
where%
$$
F(u,v,z)=1/\{\sqrt{|h_3(u,v,z)|}[s_{1(0)}\left( u,v\right) +\frac
12\int\limits_{z_0\left( u,v\right) }^z\sqrt{|h_3(u,v,z)|}dz]\},
$$
$s_{1(0)}\left( u,v\right) $ and $z_0\left( u,v\right) $ are some functions
of necessary smooth class. We note that if we put $h_4=a_4(u,v)$ the
equations (\ref{einsteq3b}) are satisfied for every $h_3=h_3(u,v,\varphi ).$

Every d--metric (\ref{rel1}) with coefficients of type (\ref{relh1h}), (\ref
{relh1}) and (\ref{relh1a}) solves the Einstein equations (\ref{einsteq3a}%
)--(\ref{einsteq3d}) with the diagonal momentum d--tensor
$$
\Upsilon _\beta ^\alpha =diag\left[ 0,0,-\varepsilon =-m_0,0\right] ,
$$
when $r_g=2\kappa m_0;$ we set the light constant $c=1.$ If we choose
$$
a_4\left( u,v\right) =\frac{\sinh ^2u\ \sin ^2v}{\sinh ^2u+\sin ^2v}
$$
our solution is conformally equivalent (if not considering the time--time
component) to the hypersurface metric (\ref{hsuf1}). The condition of
vanishing of the coefficient (\ref{relh1}) parametrizes the rotation
ellipsoid for the horizon%
$$
\frac{\widetilde{x}^2+\widetilde{y}^2}{\sigma ^2-1}+\frac{\widetilde{z}^2}{%
\sigma ^2}=\left( \frac{r_g}4\right) ^2,
$$
where the radial coordinate is redefined via relation\ $\widetilde{r}=%
\widetilde{\rho }\left( 1+\frac{r_g}{4\widetilde{\rho }}\right) ^2. $ After
multiplication on the conformal factor
$$
\left( \sinh ^2u+\sin ^2v\right) \left[ 1+\frac{r_g}{4\widetilde{\rho }}%
\right] ^4,
$$
approximating $g_1(u,v)=\cos ^2v\approx 1,$ in the limit of locally
isotropic spherical symmetry,%
$$
\widetilde{x}^2+\widetilde{y}^2+\widetilde{z}^2=r_g^2,
$$
the d--metric (\ref{rel1}) reduces to
$$
ds^2=\left[ 1+\frac{r_g}{4\widetilde{\rho }}\right] ^4\left( d\widetilde{x}%
^2+d\widetilde{y}^2+d\widetilde{z}^2\right) -\frac{\left[ 1-\frac{r_g}{4%
\widetilde{\rho }}\right] ^2}{\left[ 1+\frac{r_g}{4\widetilde{\rho }}\right]
^2}dt^2
$$
which is just the Schwazschild solution with the redefined radial coordinate
when the space component becomes conformally Euclidean.

So, the d--metric (\ref{rel1}), the coefficients of N--connection being
solutions of (\ref{einsteq3c}) and (\ref{einsteq3d}), describe a static 4D
solution of the Einstein equations when instead of a spherical symmetric
horizon one considers a locally anisotropic deformation to the hypersurface
of rotation elongated ellipsoid.

\vskip0.2cm

\paragraph{\qquad Flattened rotation ellipsoid coordinates}

${~}$
\\ ${\qquad}$ In a similar fashion we can construct a static 4D black hole
solution with the horizon parametrized by a flattened rotation ellipsoid
\cite{korn},
$$
\frac{\widetilde{x}^2+\widetilde{y}^2}{1+\sigma ^2}+\frac{\widetilde{z}^2}{%
\sigma ^2}=\widetilde{\rho }^2,
$$
where $\sigma \geq 0$ and $\sigma =\sinh u.$

The space 3D special coordinate system is defined%
$$
\widetilde{x}=\widetilde{\rho}\cosh u\sin v\cos \varphi ,\ \widetilde{y}=%
\widetilde{\rho}\cosh u\sin v\sin \varphi ,\ \widetilde{z}=\widetilde{\rho}%
\sinh u\cos v,
$$
where $0\leq u<\infty ,\ 0\leq v\leq \pi ,\ 0\leq \varphi <2\pi .$

The hypersurface metric is
\begin{eqnarray}
g_{uu} &=& g_{vv}=\widetilde{\rho}^2\left( \sinh ^2u+\cos ^2v\right) ,
 \nonumber \\
g_{\varphi \varphi } &=&\widetilde{\rho}^2\sinh ^2u\cos ^2v.
 \nonumber
\end{eqnarray}
In the rest the black hole solution is described by the same formulas as in
the previous subsection but with respect to new canonical coordinates for
flattened rotation ellipsoid.

\subsubsection{Cylindrical, Bipolar and Toroidal Configurations}

We consider a d--metric of type (\ref{dmetr4}). As a coefficient for
h--metric we choose $g_1(\chi ^1,\chi ^2)=\left( \cos \chi ^2\right) ^{2}$
which solves the Einstein equations (\ref{einsteq3a}). The energy momentum
d--tensor is chosen to be diagonal, $\Upsilon _\beta ^\alpha
=diag[0,0,-\varepsilon ,0]$ with $\varepsilon \simeq m_0=\int m_{(lin)}dl,$
where $\varepsilon _{(lin)}$ is the linear 'mass' density. The coefficient $%
h_3\left( \chi ^i,z\right) $ will be chosen in a form similar to (\ref{relh1}%
),%
$$
h_3\simeq \left[ 1-\frac{r_g}{4\widetilde{\rho }}\right] ^2/\left[ 1+\frac{%
r_g}{4\widetilde{\rho }}\right] ^6
$$
for a cylindrical elliptic horizon. We parametrize the second v--component
as $h_4=a_4(\chi ^1,\chi ^2)$ when the equations (\ref{einsteq3b}) are
satisfied for every $h_3=h_3(\chi ^1,\chi ^2,z).$

\vskip0.2cm

\paragraph{\qquad Cylindrical coordinates:}

${~}$
\\ $\qquad$
Let us construct a solution of the Einstein equation with the horizon
having the symmetry of ellipsoidal cylinder given by hypersurface formula
\cite{korn}
$$
\frac{\widetilde{x}^2}{\sigma ^2}+\frac{\widetilde{y}^2}{\sigma ^2-1}=\rho
_{*}^2,\ \widetilde{z}=\widetilde{z},
$$
where $\sigma \geq 1.$ The 3D radial coordinate $\widetilde{r}$ is to be
computed from $\widetilde{\rho }^2=\rho _{*}^2+\widetilde{z}^2.$

The 3D space coordinate system is defined%
$$
\widetilde{x}=\rho _{*}\cosh u\cos v,\ \widetilde{y}=\rho _{*}\sinh u\sin
v\sin ,\ \widetilde{z}=\widetilde{z},
$$
where $\sigma =\cosh u,\ (0\leq u<\infty ,\ 0\leq v\leq \pi ).$

The hypersurface metric is
\begin{equation}
\label{melcy}g_{uu}=g_{vv}=\rho _{*}^2\left( \sinh ^2u+\sin ^2v\right)
,g_{zz}=1.
\end{equation}

A solution of the Einstein equations with singularity on an ellipse is given
by
\begin{eqnarray}
h_3 &=&
\frac 1{\rho _{*}^2\left( \sinh ^2u+\sin ^2v\right) }\times \frac{\left[
1-\frac{r_g}{4\widetilde{\rho }}\right] ^2}{\left[ 1+\frac{r_g}{4\widetilde{\rho }}\right] ^6},  \nonumber \\
h_4 &=& a_4=\frac 1{\rho _{*}^2\left( \sinh ^2u+\sin ^2v\right) },
\nonumber
\end{eqnarray}
where $\widetilde{r}=\widetilde{\rho }\left( 1+\frac{r_g}{4\widetilde{\rho }}%
\right) ^2.$ The condition of vanishing of the time--time coefficient $h_3$
parametrizes the hypersurface equation of the horizon%
$$
\frac{\widetilde{x}^2}{\sigma ^2}+\frac{\widetilde{y}^2}{\sigma ^2-1}=\left(
\frac{\rho _{*(g)}}4\right) ^2,\ \widetilde{z}=\widetilde{z},
$$
where $\rho _{*(g)}=2\kappa m_{(lin)}.$

By multiplying the d--metric on the conformal factor
$$
\rho _{*}^2\left( \sinh ^2u+\sin ^2v\right) 
\left[ 1+\frac{r_g}{4\widetilde{\rho }}\right] ^4,
$$
where $r_g=\int \rho _{*(g)}dl$ (the integration is taken along the
ellipse), for $\rho _{*}\rightarrow 1,$ in the local isotropic limit, $\sin
v\approx 0, $ the space component transforms into (\ref{melcy}).

\vskip0.2cm

\paragraph{\qquad Bipolar coordinates:}

${~}$\\ ${\qquad}$
 Let us construct 4D solutions of the Einstein equation with the horizon
having the symmetry of the bipolar hypersurface given by the formula \cite
{korn}%
$$
\left( \sqrt{\widetilde{x}^2+\widetilde{y}^2}-\frac{\widetilde{\rho }}{\tan
\sigma }\ \right) ^2+\widetilde{z}^2=\frac{\widetilde{\rho }^2}{\sin
^2\sigma },
$$
which describes a hypersurface obtained under the rotation of the circles
$$
\left( \widetilde{y}-\frac{\widetilde{\rho }}{\tan \sigma }\right) ^2+%
\widetilde{z}^2=\frac{\widetilde{\rho }^2}{\sin ^2\sigma }
$$
around the axes $Oz$; because $|c\tan \sigma |<|\sin \sigma |^{-1},$ the
circles intersect the axes $Oz.$ The 3D space coordinate system is defined%
\begin{eqnarray}
\widetilde{x} &=&
\frac{\widetilde{\rho}\sin \sigma \cos \varphi }{\cosh \tau -\cos\sigma },
 \qquad
\widetilde{y} =
\frac{\widetilde{\rho}\sin \sigma \sin \varphi }{\cosh\tau -\cos \sigma },
 \nonumber \\
\widetilde{z} & = &\frac{\widetilde{r}\sinh \tau }{\cosh \tau
-\cos \sigma }\
\left( -\infty <\tau <\infty ,0\leq \sigma <\pi ,0\leq \varphi <2\pi \right).
 \nonumber
\end{eqnarray}
The hypersurface metric is
\begin{equation}
\label{mbipcy}g_{\tau \tau }=g_{\sigma \sigma }=\frac{\widetilde{\rho }^2}{%
\left( \cosh \tau -\cos \sigma \right) ^2},g_{\varphi \varphi }=\frac{%
\widetilde{\rho }^2\sin ^2\sigma }
{\left( \cosh \tau -\cos \sigma \right) ^2}.
\end{equation}

A solution of the Einstein equations with singularity on a circle is given
by
$$
h_3=\left[ 1-\frac{r_g}{4\widetilde{\rho }}\right] ^2/\left[ 1+\frac{r_g}{4%
\widetilde{\rho }}\right] ^6\mbox{ and }h_4=a_4=\sin ^2\sigma ,
$$
where $\widetilde{r}=\widetilde{\rho }\left( 1+\frac{r_g}{4\widetilde{\rho }}%
\right) ^2.$ The condition of vanishing of the time--time coefficient $h_3$
parametrizes the hypersurface equation of the horizon%
$$
\left( \sqrt{\widetilde{x}^2+\widetilde{y}^2}-\frac{r_g}2\ c\tan \sigma
\right) ^2+\widetilde{z}^2=\frac{r_g^2}{4\sin ^2\sigma },
$$
where $r_g=\int \rho _{*(g)}dl$ (the integration is taken along the circle),
$\rho _{*(g)}=2\kappa m_{(lin)}.$

By multiplying the d--metric on the conformal factor
\begin{equation}
\label{confbip}\frac 1{\left( \cosh \tau -\cos \sigma \right) ^2}\left[ 1+%
\frac{r_g}{4\widetilde{\rho }}\right] ^4,
\end{equation}
for $\rho _{*}\rightarrow 1,$ in the local isotropic limit, $\sin v\approx
0, $ the space component transforms into (\ref{mbipcy}).

\vskip0.2cm

\paragraph{\qquad Toroidal coordinates:}

${~}$\\ ${\qquad}$
Let us consider solutions of the Einstein equations with toroidal
symmetry of horizons. The hypersurface formula of a torus is \cite{korn}%
$$
\left( \sqrt{\widetilde{x}^2+\widetilde{y}^2}-\widetilde{\rho }\ c\tanh
\sigma \right) ^2+\widetilde{z}^2=
\frac{\widetilde{\rho }^2}{\sinh ^2\sigma }.
$$
The 3D space coordinate system is defined%
\begin{eqnarray}
\widetilde{x} &=&
\frac{\widetilde{\rho}\sinh \tau \cos \varphi }{\cosh \tau -\cos\sigma },
 \qquad
\widetilde{y} = \frac{\widetilde{\rho}\sin \sigma \sin \varphi }{\cosh
\tau -\cos \sigma }, \nonumber \\
\widetilde{z} &=& \frac{\widetilde{\rho}\sinh \sigma }{\cosh
\tau -\cos \sigma }\
\left( -\pi <\sigma <\pi ,0\leq \tau <\infty ,0\leq \varphi <2\pi \right) .
 \nonumber
\end{eqnarray}
The hypersurface metric is
\begin{equation}
\label{mtor}g_{\sigma \sigma }=g_{\tau \tau }=\frac{\widetilde{\rho }^2}{%
\left( \cosh \tau -\cos \sigma \right) ^2},
g_{\varphi \varphi }=\frac{\widetilde{\rho }^2\sin ^2\sigma }
{\left( \cosh \tau -\cos \sigma \right) ^2}.
\end{equation}

This, another type of solution of the Einstein equations with singularity on
a circle, is given by
$$
h_3=\left[ 1-\frac{r_g}{4\widetilde{\rho }}\right] ^2/\left[ 1+\frac{r_g}{4%
\widetilde{\rho }}\right] ^6\mbox{ and }h_4=a_4=\sinh ^2\sigma ,
$$
where $\widetilde{r}=\widetilde{\rho }\left( 1+\frac{r_g}{4\widetilde{\rho }}%
\right) ^2.$ The condition of vanishing of the time--time coefficient $h_3$
parametrizes the hypersurface equation of the horizon%
$$
\left( \sqrt{\widetilde{x}^2+\widetilde{y}^2}-\frac{r_g}{2\tanh \sigma }%
c\right) ^2+\widetilde{z}^2=\frac{r_g^2}{4\sinh ^2\sigma },
$$
where $r_g=\int \rho _{*(g)}dl$ (the integration is taken along the circle),
$\rho _{*(g)}=2\kappa m_{(lin)}.$

By multiplying the d--metric on the conformal factor (\ref{confbip}), for $%
\rho _{*}\rightarrow 1,$ in the local isotropic limit, $\sin v\approx 0, $
the space component transforms into (\ref{mtor}).

\subsection{A Schwarzschild like la--solution}

The d--metric of type (\ref{rel1}) is taken
\begin{equation}
\label{schla}\delta s^2=g_1(\chi ^1,\theta )d(\chi ^1)^2+d\theta
^2+h_3\left( \chi ^1,\theta ,\varphi \right) \left( \delta t\right)
^2+h_4\left( \chi ^1,\theta ,\varphi \right) \left( \delta \varphi \right)
^2,
\end{equation}
where on the horizontal subspace $\chi ^1=\rho /r_a$ is the dimensionless
radial coordinate (the constant $r_a$ will be defined below), $\chi
^2=\theta $ and in the vertical subspace $y^3=z=t$ and $y^4=\varphi .$ The
energy--momentum d--tensor is taken to be diagonal $\Upsilon _\beta ^\alpha
=diag[0,0,-\varepsilon ,0].$ The coefficient $g_1$ is chosen to be a
solution of type (\ref{btzlh4})%
$$
g_1\left( \chi ^1,\theta \right) =\cos ^2\theta .
$$
For
$$
h_4=\sin ^2\theta \mbox{ and }h_3\left( \rho \right) =-\frac{\left[
1-r_a/4\rho \right] ^2}{\left[ 1+r_a/4\rho \right] ^6},
$$
where $r=\rho \left( 1+\frac{r_g}{4\rho }\right) ^2,r^2=x^2+y^2+z^2,$ $%
r_a\dot =r_g$ is the Schwarzschild gravitational radius, the d--metric (\ref
{schla}) describes a la--solution of the Einstein equations which is
conformally equivalent, with the factor $\rho ^2\left( 1+r_g/4\rho \right)
^2,$ to the Schwarzschild solution (written in coordinates $\left( \rho
,\theta ,\varphi ,t\right) ),$ embedded into a la--background given by
non--trivial values of $q_i(\rho ,\theta ,t)$ and $n_i(\rho ,\theta ,t).$ In
the anisotropic case we can extend the solution for anisotropic (on angle $%
\theta )$ gravitational polarizations of point particles masses, $m=m\left(
\theta \right) ,$ for instance in elliptic form, when
$$
r_a\left( \theta \right) =\frac{r_g}{\left( 1+e\cos \theta \right) }
$$
induces an ellipsoidal dependence on $\theta $ of  the radial coordinate,%
$$
\rho =\frac{r_g}{4\left( 1+e\cos \theta \right) }.
$$
We can also consider arbitrary solutions with $r_a=r_a\left( \theta ,t\right)
$ of oscillation type, $r_a\simeq \sin \left( \omega _1t\right) ,$ or
modelling the mass evaporation, $r_a\simeq \exp [-st],s=const>0.$

So, fixing a physical solution for $h_3(\rho ,\theta ,t),$ for instance,
$$
h_3(\rho ,\theta ,t)=-\frac{\left[ 1-r_a\exp [-st]/4\rho \left( 1+e\cos
\theta \right) \right] ^2}{\left[ 1+r_a\exp [-st]/4\rho \left( 1+e\cos
\theta \right) \right] ^6},
$$
where $e=const<1,$ and computing the values of $q_i(\rho ,\theta ,t)$ and $%
n_i(\rho ,\theta ,t)$ from (\ref{einsteq3c}\NEG ) and (\ref{einsteq3d}),
corresponding to given $h_3$ and $h_4,$ we obtain a la--generalization of
the Schwarzschild metric.

We note that fixing this type of anisotropy,  in the locally isotropic limit
we obtain not just the Schwarzschild metric but a conformally transformed
one, multiplied on the factor $1/\rho ^2\left( 1+r_g/4\rho \right) ^4.$

\section{Final remarks}

We have presented new classes of three and four dimensional black hole
solutions with local anisotropy which are given both with respect to a
coordinate basis or to an anholonomic frame defined by a N--connection
structure. We proved that for a corresponding ansatz such type of solutions
can be imbedded into the usual (three or four dimensional) Einstein gravity.
It was demonstrated that in general relativity there are admitted static,
but anisotropic (with nonspheric symmetry), and elliptic oscillating in time
black hole like configurations with horizons of events being elliptic (in
three dimensions) and rotation ellipsoidal, elliptic cylinder, toroidal and
another type of closed hypersurfaces or cylinders.

From the results obtained, it appears that the components of metrics with
generic local anisotropy are somehow undetermined from field equations if
the type of symmetry and a correspondence with locally isotropic limits are
not imposed. This is the consequence of the fact that in general relativity
only a part of components of the metric field (six from ten in four
dimensions and three from six in three dimensions) can be treated as
dynamical variables. This is caused by the Bianchi identities which hold on
(pseudo) Riemannian spaces. The rest of components of metric should be
defined from some symmetry prescriptions on the type of locally anisotropic
solutions and corresponding anholonomic frames and, if existing,
compatibility with the locally isotropic limits when some physically
motivated coordinate and/or boundary conditions are enough to state and
solve the Cauchy problem.

Some of the problems discussed so far might be solved by considering
theories containing non--trivial torsion fields like metric--affine and
gauge gravity and for so--called generalized Finsler--Kaluza--Klein models.
More general solutions connected with locally anisotropic low energy limits
in string/M--theory and supergravity could be also generated by applying the
method of computation with respect to anholonomic (super) frames adapted to
a N--connection structure. This topic is currently under study.

\subsection*{Acknowledgments}

The author would like to thank P J Steinhardt, N G Turok and V A Rubakov for
support of his participation at NATO ASI ''Structure Formation in the
Universe'', July 26 -- August 6, 1999 (held at Isaac Newton Institute for
Mathematical Sciences, Cambridge University, UK), where this work was
communicated. He is also very grateful to H Dehnen for hospitality and
useful discussions during his visit at Konstantz University, Germany.

\vspace*{0.25cm} 

\end{document}